# Lessons from Copper Indium Gallium Sulfo-Selenide Solar Cells for progressing Thin Film Perovskite Photovoltaic Technology

Authors: Mirjana Dimitrievska[1*], Edgardo Saucedo[2,3], Stefaan De Wolf,[4] Billy J. Stanbery[5,6], Veronica Bermudez Benito[7*]

Affiliations:

1 Nanomaterials Spectroscopy and Imaging Group, Transport at Nanoscale Interfaces Laboratory, Swiss Federal Laboratories for Materials Science and Technology (EMPA), Überlandstrasse 129, 8600 Dübendorf, Switzerland

2 Universitat Politècnica de Catalunya (UPC), Micro and Nanotechnology Group, Thin Film Photovoltaics Laboratory, Escola d'Enginyeria de Barcelona Est (EEBE), Av. Eduard Maristany 10-14, Barcelona 08019, Catalonia, Spain

3 Universitat Politècnica de Catalunya (UPC), Barcelona Centre for Multiscale Science & Engineering, Av. Eduard Maristany 10-14, Barcelona 08019, Catalonia, Spain

4 Center for Renewable Energy and Storage Technologies (CREST), Physical Sciences and Engineering Division, King Abdullah University of Science and Technology, Thuwal 23955-6900, Saudi Arabia

5 Clean Energy Institute, Department of Chemical Engineering, University of Washington, 3781 Okanogan Lane, Seattle, WA 98195-1750, United States of America

6 Department of Biological and Chemical Engineering, Colorado School of Mines, 920 15[th] Street, Hill Hall 201, Golden, CO 80401, United States of America

7 Berbetin, 505 Chemin de Rabiac, 06600 Antibes, France

*Corresponding Author: mirjana.dimitrievska@empa.ch, vbermudezbenito@berbetin.com

**Abstract**

The growing demand for photovoltaic (PV) technologies that are lightweight, flexible, and can be seamlessly integrated into diverse applications has propelled interest in thin-film solar cells. Among these, Cu(In,Ga)(S,Se)$_2$ (CIGS) and metal halide perovskites have garnered significant attention in the past and present, respectively. While CIGS reached commercial readiness after decades of refinement, their large-scale deployment was hindered by manufacturing complexity, scale-up challenges, and a lack of coordination between materials, device design, and production systems. Perovskite solar cells, despite setting record efficiencies at an unprecedented pace, now face similar challenges on their path to commercialization: ensuring long-term stability, translating laboratory performance to scalable architectures, and aligning with industrial realities. In this perspective, we revisit the CIGS experience not as a benchmark, but as a blueprint, highlighting how its successes and failures can inform a more deliberate and durable trajectory for perovskite PV. By bridging this historical perspective with the current frontier, we propose that the future of perovskites depends not only on continued innovation, but also on learning from past thin-film PV experiences to avoid     repeating their pitfalls.

1. **Introduction**

The global transition toward an all-electrified society will demand photovoltaic (PV) technologies that are not only efficient and durable but also lightweight, versatile, and manufacturable at scale and close to their point of use. While crystalline silicon currently dominates the PV market due to its mature supply chains and long-term reliability, it faces limitations in applications requiring high power-to-weight ratios, flexible form factors, or seamless integration into mobile or structural platforms. Thin-film technologies—including , the long-studied chalcopyrite-based copper indium gallium selenide and sulfide (Cu(In,Ga)(S,Se)$_2$, or CIGS) and the emerging metal halide perovskites (MHPs)—may offer distinct advantages in such contexts. The low thickness of their absorber, tunable optoelectronic properties, and compatibility with lightweight or flexible substrates position them as complementary to silicon rather than direct competitors.

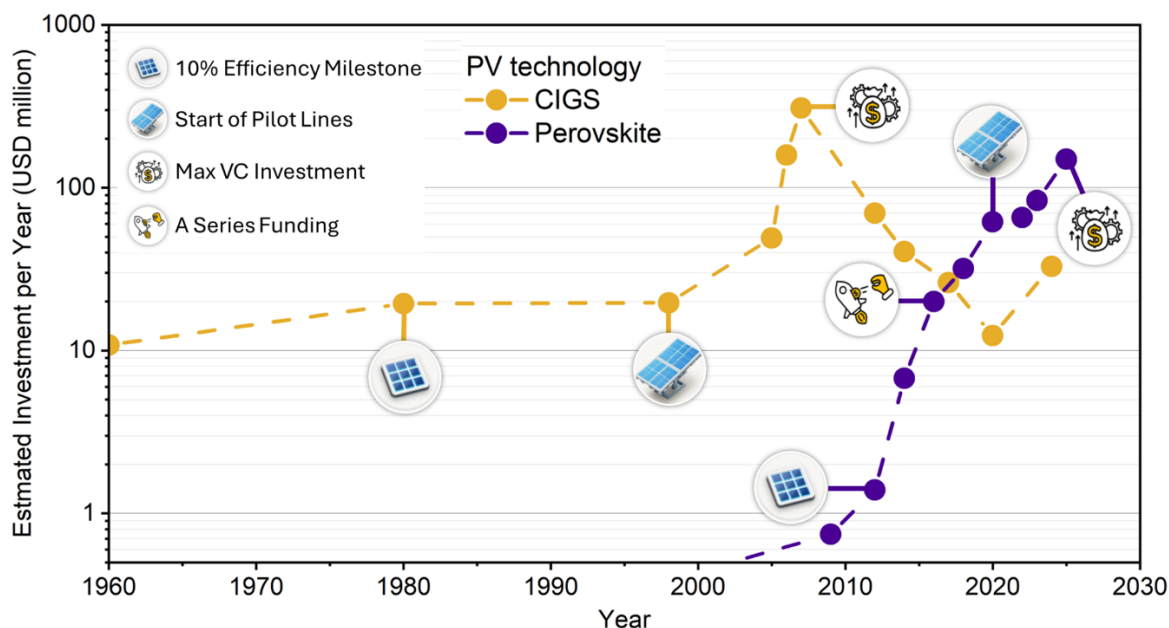

**Key Highlights**

**CIGS PV Technology:**

**1980s**: CuInSe$_2$ thin-film lab cells reach >10% efficiency at Boeing, NREL, and Siemens labs
**1998**: Shell Solar and Siemens start CIGS pilot lines (~MW scale)
**2006**: Nanosolar raises ~$100M (~$150M in 2025) for roll-to-roll CIGS printing pilot line
**2007**: Peak venture capital interest in CIGS; HelioVolt, Solyndra raise >$150M combined
**2012**: Post-financial crisis: Nanosolar, Solyndra, HelioVolt collapse
**2020**: Japan's Solar Frontier remains the only large-scale CIGS manufacturer
**2024**: Resurgence in R&D and commercial activity — tandems, BIPV, new factories (Avancis, Midsummer, Roltec)

**Perovskite PV Technology:**

**2009**: Miyasaka et al. report first perovskite solar cell (~3.8% efficiency)
**2012**: Snaith's group introduces solid-state perovskite solar cells; 10% efficiency milestone
**2014**: Oxford PV, Saule Tech, Solaronix raise seed funding
**2016**: Start of serious Series A funding rounds; EU and US DOE back early commercialization programs
**2018**: Efficiency record exceeds 23%; Tandem R&D backed by EU's Horizon 2020 and Japan's NEDO programs
**2020**: Oxford PV raises ~£65M (~£75M in 2025); Tandem PV (US), UtmoLight (China), and others begin pilot projects
**2023**: UtmoLight (China), Wonder Solar, and Oxford PV announce production expansion plans
**2025**: Cumulative global investment from venture capital, national programs, and GWh-scale fabs crosses $500M+

**Figure 1. Global investment trends and key historical milestones in CIGS, and Perovskite photovoltaic (PV) technologies from 1960 to 2025.** The log-scale plot displays estimated annual investments (USD million per year adjusted for inflation) for each technology, based on key funding events, public disclosures, and institutional reports. The lines serve as visual guides to illustrate investment trajectories over time, rather than representing continuous or cumulative investment. Custom milestone icons highlight major events in PV development, including the first attainment of 10% power conversion efficiency, the launch of pilot production lines, peak venture capital funding, and cost breakthroughs such

as module prices dropping below $1/Wp. The annotated timeline contextualizes these inflection points across R&D, commercialization, and manufacturing scale-up phases.

Interest in CIGS photovoltaic materials has been driven by the search for higher performance, broader tunability, and reduced toxicity concerns. CIGS, as a more complex multinary absorber, offered the potential for bandgap engineering, higher theoretical power conversion efficiencies (PCEs), and a more favorable environmental profile compared to CdTe technology due to the absence of cadmium, prompting intense research and commercial exploration starting in the 1990s, further stimulated by the high cost of silicon wafers at that time.

CIGS solar cells are among the most mature thin-film technologies, with decades of research culminating in lab-scale device PCEs exceeding 23% and commercial module production.[2,3] However, despite their strong technical potential, the commercial success of CIGS has been limited. A wave of companies emerged between the late 1990s and early 2010s—including Solar Frontier, Global Solar Energy, HelioVolt, MiaSolé, Solibro, Flisom, Odersun, Nanosolar, and Solyndra—driven by venture capital and the promise of thin-film alternatives to silicon. Yet many of these companies struggled to advance beyond prototyping or pilot production. The challenges were multifaceted: CIGS manufacturing required significant capital investment, driven in part by the complexity of multi-element co-deposition, which demands tight control over stoichiometry and phase formation. While early concerns focused on overcoming limited deposition rates and potential scaling issues—particularly due to selenium vapor scattering and homogeneity over large substrate areas—subsequent studies have shown these were manageable with optimized process control and source design.[4] More critically, the sharp decline in crystalline silicon wafer costs and cell performance improvements post-2011 eroded CIGS's cost advantage before it could mature industrially. Today, only several companies remain active in CIGS manufacturing, notably Avancis/CNBM (China & Korea), Midsummer (Italy & Sweden), and Roltec (Poland), with capacities from tens to hundreds of MW. This outcome serves as a cautionary tale about the difficulties of scaling promising laboratory technologies without aligning them with industrial and market realities.

This history may offer valuable lessons for emerging PV contenders like perovskite solar cells, of which its rise towards commercialization has been perceived by several thin-film PV veterans as a 'déjà vu'. Perovskite based solar devices have rapidly advanced in laboratory performance, with PCEs recently surpassing 26% and 34% in single-junction and perovskite/silicon tandem configurations, respectively.[5] Indeed, their low-temperature processability, either via solution or vacuum deposition techniques (or a combination thereof), excellent opto-eletronic properties and compositional tunability offer compelling opportunities for tandem integration with silicon, extending further the performance trajectory of mainstream PV, or for deployment in specialized applications—such as high-altitude platforms, drones, satellites or building-integrated PV—where conventional silicon modules may fall short.[6] However, in either scenario, perovskites still face fundamental barriers to commercialization, including homogeneity challenges with size increase, long-term stability, environmental sensitivity, and the need for industrially viable architectures.

Several startups have emerged to industrialize perovskite PV, including (but not limited to) Oxford PV, Caelux, CubicPV, Saule Technologies, Swift Solar, Tandem PV, GCL Optoelectronics, Renshine and Microquanta; many of the major silicon companies (such as Hanwa Q-Cells, Longi, Jinko, Trina, Tongwei, among others) also pursue very actively perovskite research, the latter usually with an emphasis on perovskite/silicon tandems. These ventures have achieved remarkable lab-scale milestones—such as certified perovskite/silicon tandem PCEs approaching 35% (Longi)—and initiated pilot production lines.[7] Yet despite over a decade of R&D and significant capital investment, no perovskite solar module has reached true commercial availability, even though Oxford PV shipped its first perovskite–silicon tandem modules in 2024. This delay is primarily due to unresolved challenges in pairing high performance with long-term operational stability under real-world conditions and reliable scaling techniques from small cells to large-area modules. The absence of a unified/standardized manufacturing roadmap and the difficulty of translating lab-scale processes into industrial reproducibility and throughput continue to constrain perovskite deployment.

Figure 1 summarizes the annual investment trajectories and key historical milestones for CIGS and perovskite PV technologies between 1960 and 2025. The data points represent estimated yearly investments (in USD million per year), based on public records, company disclosures, and major program announcements, with the plotted lines serving as guides to visualize overall trends rather than continuous data. Superimposed icons mark key inflection points—such as efficiency breakthroughs, pilot line launches, venture capital peaks, and manufacturing cost thresholds—that shaped each technology's industrialization path. These annotated timelines contextualize the relative pace and maturity of each thin-film technology and frame the comparative analysis that follows.

In this perspective, we provide a comparative analysis of CIGS and perovskite PV technologies across their material properties, device architectures, defect landscapes, degradation mechanisms, and manufacturing challenges. Rather than framing these technologies in opposition, we explore how the historical trajectory of CIGS can inform the path forward for perovskites—particularly in avoiding missteps related to long-term reliability, scalability and supply chain constraints. Simultaneously, we examine how the rapid innovation cycles of perovskites might inspire renewed directions for thin-film PV more broadly. Our aim is to identify the transferable lessons and critical differentiators that can guide the development of next-generation PV technologies capable of serving both conventional markets and emerging application spaces.

## 2. Comparative Framework for Materials, Devices and Reliability

### 2.1. Fundamental Material Properties

Chalcopyrites such as $Cu(In,Ga)(S,Se)_2$ (CIGS) and MHPs such as the archetypical $MAPbI_3$ composition exemplify two contrasting paradigms in PV absorber materials. Chalcopyrites are defined by their structurally rigid, covalent-ionic lattice with high thermal and chemical stability.[8] Perovskites, on the other hand, offer a softer, highly ionic framework that allows for low-temperature processing and feature a high level of defect tolerance.[9,10] These foundational differences manifest in distinct defect physics: CIGS exhibits a complex landscape of native point defects and defect clusters, often requiring precise control over composition and process conditions to suppress deep-level traps.[11] In contrast, MHPs typically only host shallower defects with benign electronic consequences, enabling efficient devices using solution-processed fabrication.

Ion migration provides a critical divergence. CIGS metastability is often linked to Cu/Na migration, but alternative explanations such as lattice relaxation around Se vacancies have also been proposed. These effects are generally reversible and stabilize over time.[12,13] MHPs, however, exhibit rapid migration of halide and organic ions, which underpins phenomena such as hysteresis, phase segregation, long-term degradation or even catastrophic failure.[14,15] Grain boundaries in CIGS have been shown to be electronically ambivalent with some reports suggesting possible benefits but without clear evidence for efficiency gains, , whereas in MHPs they are typically considered electronically benign but remain ion migration pathways.[16-18] These contrasts highlight the importance of mechanical rigidity and defect control in determining long-term performance.

### 2.2. Device Architectures and Interfaces

These differences in intrinsic properties shape specific device integration strategies. CIGS solar cells have matured around a substrate configuration with an opaque Mo back contact (collecting holes), transparent CdS or Zn(O,S) buffer layers (collecting electrons), and well-optimized heterojunctions. These exclusively p-n polarity architectures, while less flexible, have demonstrated remarkable longevity in field-deployed modules. Its constraint lies in the limited range of compatible contact materials due to high doping levels and compositional sensitivity.

MHPs, by contrast, support a wide array of device architectures—planar and mesoporous, n–i–p and p–i–n—enabled by low-temperature processing, ambipolar transport, and a relatively wide variety of charge selective materials.[19,20] This versatility has accelerated early performance gains but also complicates scalability and reproducibility, as the lack of standardized architectures makes large-area, reproducible manufacturing more difficult. Charge transport layers in perovskite cells range from sensitive molecular semiconductors, often fullerenes (to collect electrons) and self-assembled monolayers (to collect holes), to inorganic alternatives like $TiO_x$, $SnO_x$ and $NiO_x$.[21,22] Interfaces in MHPs are far more unstable than in CIGS, often susceptible to interdiffusion, ion accumulation, or chemical degradation. Efforts in perovskites increasingly mirror lessons from CIGS: the importance of interfacial passivation, chemical compatibility, and architectural simplification.[3,23,24]

## 2.3. In Operando Behavior

Device behavior under operational stress further underscores these differences. CIGS exhibits metastabilities driven by defect redistribution but these effects have been found to stabilize over time.[25,26] The performance variability tends to be gradual and manageable, aided by decades of empirical understanding, process sophistication and field validation.

In contrast, perovskites exhibit a more dynamic and less predictable response under operational conditions with multiple underlying degradation mechanisms. For instance, ion migration interacts with electric fields and interfaces, altering device performance through hysteresis, field screening, and in some cases, irreversible chemical change.[27,28] Whereas record performance certification by accredited labs is common, these instabilities pose a barrier to independent verification of stability claims, which would be desirable towards overall research progress as well as investor confidence. CIGS experience suggests that early emphasis on standardized stability testing, combined with in-line diagnostics during fabrication, are critical for mitigating such effects.

## 2.4. Environmental Stability

Perhaps the most consequential divergence lies in environmental durability. CIGS modules, backed by long-term field data, have proven to be resilient to moisture, oxygen, and thermal cycling, with degradation often confined to unsealed module edges or poorly laminated regions. In contrast, MHPs remain highly susceptible to hydrolysis, photo-oxidation, and thermal stress, including contact delamination. Ion migration exacerbates this vulnerability by accelerating degradation pathways.[29]

Improving perovskite stability has led to strategies familiar to the CIGS community: alloy or compositional engineering, introduction of barrier layers, and adequate, robust encapsulation.[30,31] However, perovskites will require deeper integration of environmental mitigation at the materials design stage. The CIGS trajectory illustrates that stability cannot be retrofitted—it must be engineered from the outset.

Taken together, the comparison of these two thin-film materials reveals not just a difference in chemistry or architecture, but in technological philosophy, as illustrated in Figure 2. Where CIGS matured through incremental optimization, standardization, and field validation, perovskites face the challenge of translating laboratory excellence, often driven by academic validation, into durable, scalable technology with very limited field operational knowledge to date. Lessons from CIGS offer both a warning and a roadmap for the next phase of perovskite PV development.

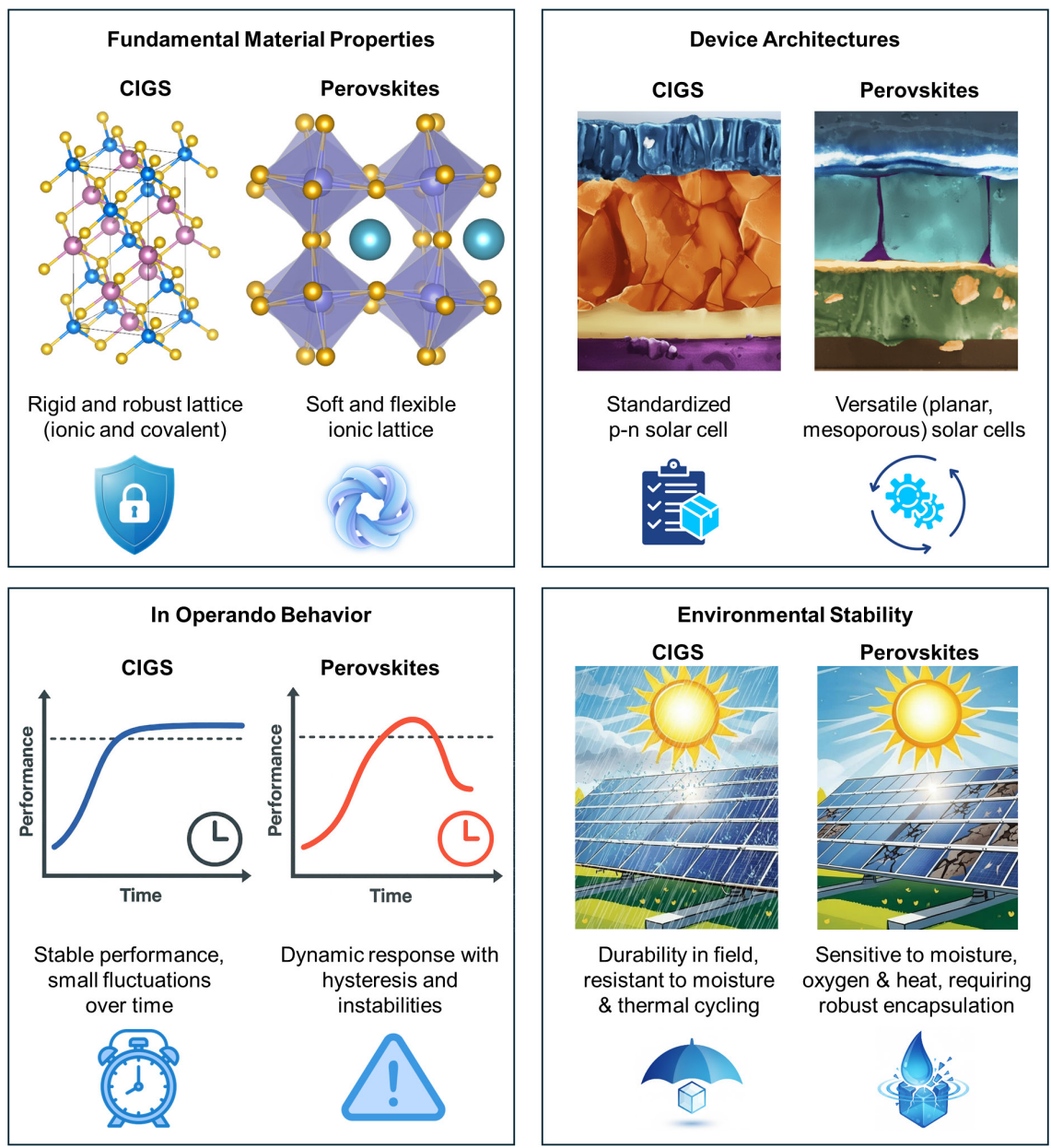

**Figure 2.** Comparative framework for CIGS and perovskite photovoltaics. Schematic overview of key contrasts between CIGS and perovskite solar cells in terms of (top left) fundamental material properties, (top right) device architectures, (bottom left) in operando behavior, and (bottom right) environmental stability. CIGS exhibits a rigid lattice, standardized p–n device structure, stable field performance, and long-term durability, while perovskites feature a soft ionic lattice, versatile device architectures, dynamic operational response, and higher environmental sensitivity requiring robust encapsulation.

## 3. Towards industrialization, reliability and durability

### 3.1. Manufacturing/Scale-Up

The manufacturing trajectories of CIGS and MHPs reflect contrasting philosophies in process control, scalability, and industrial readiness. CIGS development prioritized long-term stability and reproducibility, but this came at the expense of process complexity and inability to fully translate lab scale records to practical module size[32]. Commercial production of CIGS relies primarily on vacuum-based deposition methods such as co-evaporation and sputtering, which offer precise compositional and thickness control but suffer from deposition rate constraints and demand careful tuning and monitoring of stoichiometry, substrate temperature, and phase formation. Although CIGS is often described as compositionally tolerant, this flexibility is limited: off-stoichiometric growth outside narrow process windows can result in detrimental secondary phases, interfacial instability, and electronic defects.[3,33] Accordingly, post-deposition treatments—including alkali doping (e.g., Na, K, Rb) and annealing—are essential to achieve high-efficiency devices with controlled carrier concentrations and minimal recombination.[34]

In contrast, the scale-up of perovskite PV has been driven by rapid device performance gains mainly under solution-processed, low-temperature conditions. Techniques such as slot-die coating, blade coating, and inkjet printing offer pathways toward large-area fabrication, but achieving uniform and defect-free films over large substrates remains a major challenge. Most record efficiencies still come from spin-coating on small areas, whereas scalable methods like slot-die or blade coating perform lower due to solvent and crystallization dynamics. Vacuum-based evaporation methods are increasingly explored to improve control over film composition, morphology, and interfacial quality.[35,36] However, scalable manufacturing of MHPs still faces persistent challenges, including film uniformity, phase segregation, sensitivity to ambient conditions, batch-to-batch reproducibility, sufficiently high deposition rates and overall throughput. Compounding these issues is the diversity of device architectures in the field, ranging from n–i–p to p–i–n stacks with widely varying contact layers and transport materials. This architectural heterogeneity complicates efforts toward process standardization and slows industrial convergence, yet also fosters cross-fertilization of ideas and materials, potentially accelerating innovation across different device configurations.

Yet, while CIGS was less prone to degradation than MHPs and benefited from decades of iterative refinement, its commercial impact remained limited. Manufacturing complexity, limited throughput, and inconsistent production quality prevented many ventures from achieving cost-effective, large-scale deployment. Several well-positioned companies—including Solar Frontier, HelioVolt, and MiaSolé— struggled to meet long-term performance and scalability targets. Others, such as Nanosolar and Solyndra, failed due to unresolved issues in new technology development, materials control, ink formulation, process reproducibility, and annealing stability, leading to significant gaps between laboratory promise and manufacturing reality. As investor confidence waned, the situation was further compounded by the rapid cost declines and performance improvements of crystalline silicon, which diverted capital and attention away from CIGS. [37,38]

These historical outcomes offer important cautionary lessons for the commercialization of perovskite PV. A narrow focus on achieving record efficiencies—without simultaneously addressing scalability, reproducibility, and environmental durability—risks reproducing the same systemic misalignments that constrained CIGS. To avoid this trajectory, perovskite development must move beyond conventional technology maturity frameworks and embed manufacturability considerations into the research pipeline from the outset. This includes integrating stability engineering, supply chain foresight, and in-line process control directly into materials and device development.

Crucially, perovskite research must embrace Scaling Readiness Level (SRL) frameworks alongside the more established Technology Readiness Level (TRL) metrics[39]. While TRL tracks functional milestones—such as device efficiency and stability under lab conditions—SRL gauges the practicality of scaling a process using industrially relevant tools, materials, and environments, to bridge TRL to deployment scale. Many perovskite technologies now claim TRL 6 or higher, yet still operate at a SRL of 2–3, relying on low throughput methods in nitrogen-filled gloveboxes and the use of expensive lab-grade precursors. These workflows, while suitable for academic studies, provide no realistic path toward real-world manufacturing. Recognizing and closing the TRL–SRL gap is essential if perovskite PV is to achieve not only technical breakthroughs but also economic viability and manufacturing convergence.

Beyond caution, the CIGS experience offers a valuable technical and strategic foundation for perovskite manufacturing. Decades of expertise in multilayer deposition, interface passivation, and lifetime reliability testing can be adapted to MHPs, especially as vacuum-based fabrication methods gain traction in perovskite processing.[40] The reliability protocols developed for CIGS—such as accelerated aging under light, heat, humidity, and voltage stress—provide a ready-made framework for benchmarking perovskite device stability.[41]

CIGS' challenge of balancing innovation with reproducibility underscores that success depends as much on process compatibility and standardization as on materials performance. For perovskites to advance, innovation must be paired with clear market focus and disciplined development—anchored in stable compositions, compatible architectures, reproducible fabrication, and scalable manufacturing strategies that deliver efficiency and long-term durability while meeting industrial metrics of throughput, yield, and cost-per-watt.[42]

To meet these goals, next-generation perovskite materials must combine thermal and photochemical stability for long-term outdoor exposure with low ion mobility to suppress hysteresis, phase segregation, and interfacial instability. They should display compositional tolerance to enable reproducible large-area fabrication, as well as resilience against moisture and oxygen to reduce reliance on encapsulation. Non-toxic or mitigated-toxicity chemistries will ease environmental regulation, while compatibility with scalable transport layers and contacts can avoid unstable or costly molecular semiconductors. Finally, stable interfaces under bias and illumination are essential to prevent degradation at heterojunctions and charge extraction layers.

Embedding these materials traits into the **design-for-manufacturing mindset** from the beginning will be critical. In this context, the CIGS experience is not merely a historical case study—it is a strategic playbook for how to evolve from scientific discovery to robust, manufacturable technology.

### 3.2. Packaging and Durability

Packaging and durability are central to PV viability. For CIGS, decades of field use show performance retention over 25 years, supported by robust encapsulation and intrinsic absorber stability. Although CIGS surfaces are sensitive to oxidation and photodegradation, these effects are largely suppressed in devices through encapsulation and the CdS buffer layer. The buffer serves multiple roles: forming a chemically compatible junction with the p-type absorber, passivating surface states, ensuring favorable band alignment, and stabilizing the interface under operation and exposure. This differs from perovskites, where buffer layers such as ALD $SnO_x$ or $MoO_x$ mainly protect sensitive layers during fabrication. CIGS degradation follows predictable, thermally activated, moisture-accelerated pathways described by Arrhenius kinetics, enabling lifetime extrapolation. Standardized qualification protocols are well established, with industrial modules passing IEC tests and showing <1% annual field degradation.

For MHPs, stability remains less mature. Despite advances in intrinsic and interfacial robustness, no commercially sized modules with proven long-term durability have been shipped. Perovskites remain highly vulnerable to oxygen, heat, UV light, and humidity, leading to degradation, phase segregation, and ion migration; laser scribing also raises reliability concerns. Encapsulation strategies such as glass–glass lamination, edge sealing, and multilayer barriers show promise but require validation at module scale and under field stress. Importantly, encapsulation alone is insufficient—materials engineering, defect passivation, and contact stability are equally critical.[43]

The durability trajectory of CIGS provides both a benchmark and a roadmap for perovskite PV. It underscores the need for early, standardized testing under outdoor stressors and for co-designing encapsulation with absorber and contact stability. CIGS progress was enabled by feedback loops between field data and materials development, a systems-level approach now emerging for perovskites. Promising initiatives—such as PACT (U.S.), HZB (Germany), EPFL–CSEM (Switzerland), and KAUST (Saudi Arabia)—are beginning to generate long-term field data, though such datasets remain scarce. Real-world applications will require advanced encapsulants and deeper insight into degradation mechanisms, with CIGS offering lessons for reliable perovskite deployment.

### 3.3. Device Reliability During In-Operando Field Conditions

The long operational track record of CIGS PV offers a valuable framework for understanding—and anticipating—the reliability challenges facing perovskite solar cells. Decades of CIGS field deployment across diverse climates have resulted in a comprehensive map of degradation modes and mitigation strategies. These include thermally activated recombination losses, reversible light-soaking effects, and moisture-induced charge-selective contacts/absorber reactions—all well-characterized through standardized testing and long-term outdoor monitoring.[44] This foundation enabled robust lifetime modeling and IEC qualification, ultimately supporting product warranties of 25 years or more. For perovskites, which are still in the pre-commercial stage, learning from this systematic reliability engineering is crucial to bridge the gap between lab stability and field robustness.

One of the most instructive parallels lies in metastable behavior under illumination. In CIGS, light soaking can transiently improve the open circuit voltage ($V_{OC}$) through electric-field redistribution linked to defect migration, particularly copper and sodium ions.[3,45] These effects are now understood to be largely reversible, and device stabilization protocols have been implemented in manufacturing. In perovskites, however, ion migration occurs at much faster rates and involves multiple species—including halides and organic cations—that destabilize internal electric fields, induce hysteresis, and create local phase segregation. CIGS teaches that even "reversible" ion motion must be managed through materials and interface design, not merely through encapsulation. The use of alkali dopants in CIGS, once seen as a minor adjustment, evolved into a key reliability lever—suggesting that strategic dopant engineering in perovskites may hold similar potential.

Grain boundary behavior is another domain where CIGS experience offers sobering guidance. Early studies suggested beneficial grain boundary effects in CIGS, but later work and simulations showed that enhanced current collection at grain edges does not necessarily translate to improved device efficiency, particularly in wide-bandgap compositions.[46] Kelvin Probe Force Microscopy revealed that apparent band bending at CIGS grain boundaries can be artifact-driven, depending on imaging mode and topography.[47] Similarly, perovskites show ion accumulation, recombination, and photodegradation at grain edges[48,49] underscoring that device stability is limited by weak domains and highlighting the value of advanced spatial diagnostics like luminescence mapping and Electron Beam–Induced Current (EBIC).

CIGS modules are thermally robust, with absorbers stable above 200 °C, though device stability is often limited to ~120 °C by junction and interface degradation. Failures typically arise from encapsulation or delamination, and degradation is predictable and largely decoupled from absorber chemistry. Perovskites, by contrast, undergo phase transitions, organic decomposition, and interfacial reactions at much lower thresholds (~85 °C), especially under bias. Lessons from CIGS—minimizing thermal gradients, matching expansion coefficients, and adopting passive cooling or UV filtering—are critical for stable perovskite tandems and flexible devices.

One of the most important lessons from CIGS is the role of outdoor validation. Long-term exposure of commercial modules revealed degradation not evident in accelerated indoor tests—such as moisture ingress at edge seals, seasonal drift, and potential-induced degradation (PID) under system voltage.[50] These insights directly informed CIGS module redesign and manufacturing. By contrast, perovskite development relies mainly on accelerated ISOS testing with limited field data, highlighting the need for early outdoor trials—even at sub-commercial scale—to identify stressors, test encapsulation, and validate stability.

Finally, CIGS demonstrates that achieving reliability is not just about materials resilience, but about system-level co-design. Encapsulation was not an afterthought in CIGS—it evolved alongside absorber engineering, contact optimization, and mechanical stability. In perovskites, similar systems thinking is needed. Encapsulation, charge transport layer chemistry, and even module architecture must be considered in concert. The fragmented approach seen in perovskite R&D—where absorbers, interfaces, and packaging are often developed independently—may delay convergence on reliable, manufacturable products. The CIGS trajectory shows that reliability is not achieved through heroic materials alone, but through deliberate coordination of every layer, interface, and environmental interaction.

As highlighted in Figure 3, CIGS shows that reliability is a systems property built through co-design, standardization, and feedback from field data. For perovskites, this means moving beyond efficiency records to address scalability, durability, and industrial compatibility from the outset. Bridging the TRL–SRL gap, validating stability in outdoor pilots, and simplifying device stacks will be critical. Early adoption of this discipline can help perovskites avoid CIGS' pitfalls and progress toward durable, manufacturable solar technologies.

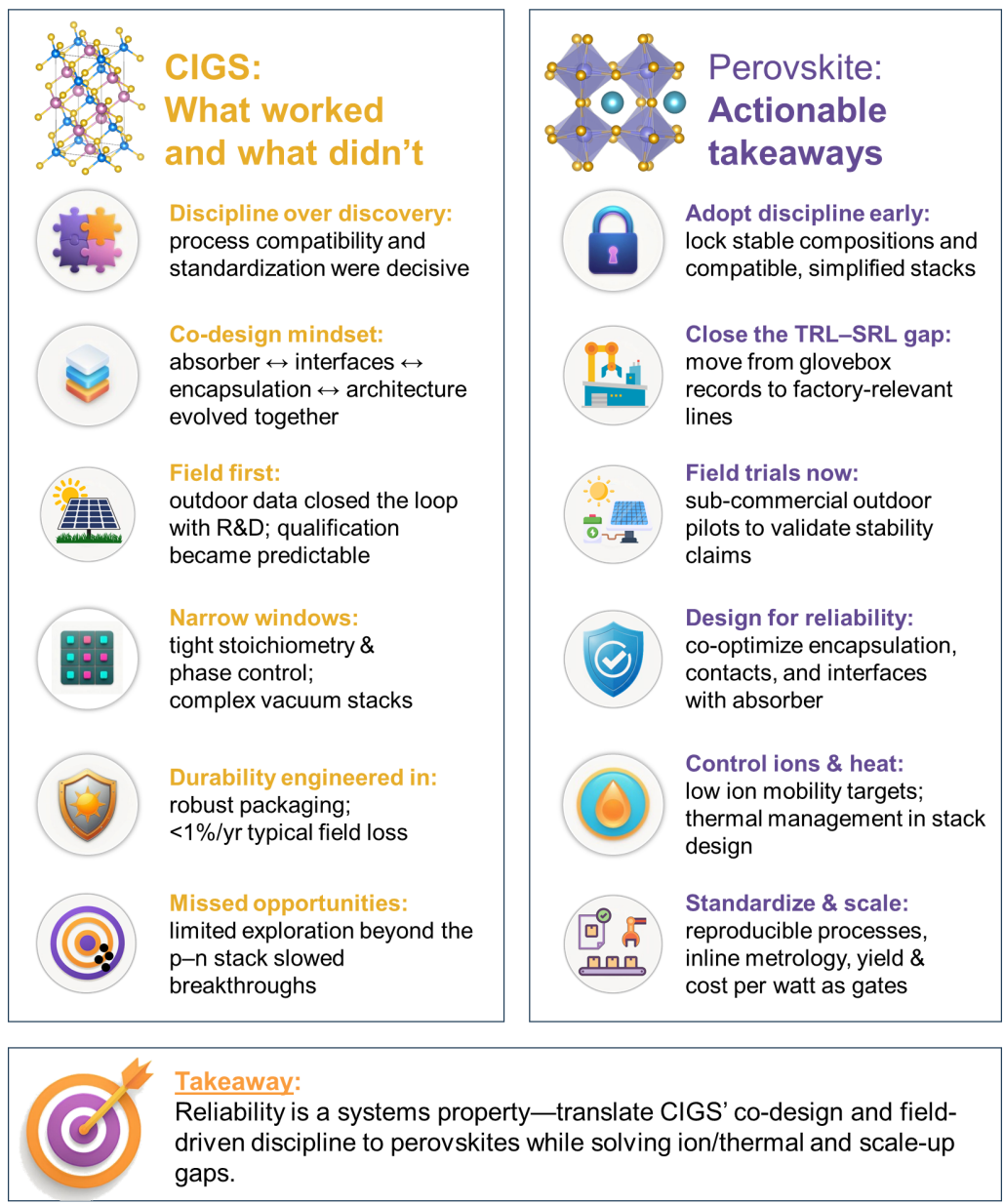

**Figure 3. System-level lessons from CIGS for perovskite PV** This two-panel infographic contrasts the CIGS experience with actionable directions for perovskites. On the left, CIGS highlights the importance of process compatibility, standardization, and co-design across absorber, interfaces, encapsulation, and device

architecture—supported by outdoor validation and robust packaging that enabled predictable reliability, but constrained by narrow process windows and limited architectural exploration. On the right, the perovskite pathway emphasizes the need to adopt discipline early with stable, simplified stacks, bridge the TRL–SRL gap, establish field trials, design reliability into the stack, manage ion and thermal effects, and standardize scalable processes aligned with industrial metrics.

## 4. Conclusions

Perovskite photovoltaics represent a transformative opportunity for thin-film solar energy, combining tunable optoelectronic properties with low-temperature processing and high efficiencies. Yet despite remarkable laboratory progress, perovskites have yet to cross the threshold into stable, field-deployed modules. In this regard, the development arc of CIGS solar cells offers more than historical context—it provides a strategic lens through which to anticipate and address the technological and industrial hurdles facing perovskites today.

CIGS taught us that stability, manufacturability, and system integration must be addressed in parallel with and inform materials innovation. Its path from laboratory cells to commercial modules was shaped by hard-won lessons in compositional control, interfacial reliability, long-term testing, and industrial convergence. Many of its challenges—such as managing ion migration, encapsulating reactive materials, and standardizing processing—are now resurfacing in the perovskite field, often with higher urgency due to the inherent mobility of hybrid halide species and greater architectural diversity. The collapse of CIGS manufacturers further underscores that materials promise alone is not sufficient: robust supply chains, knowledge of the rapidly evolving PV landscape, process standardization, strong processing quality control and demonstrable outdoor reliability must co-evolve with performance gains.

For perovskites to mature into a viable photovoltaic technology, they must move beyond laboratory metrics and adopt the discipline of durable design. This includes establishing stability protocols, deploying outdoor demonstrators, and embedding scalability in every layer of the device stack. The CIGS experience offers a clear precedent: success requires not just innovation, but patience, integration, and a commitment to learning from the field. If embraced fully, these lessons could help perovskite photovoltaics succeed where previous thin-film technologies fell short—not by replacing them, but by evolving more wisely.